\documentclass[twocolumn,showpacs,preprintnumbers,amsmath,amssymb]{revtex4}

\usepackage{graphicx}
\usepackage{dcolumn}
\usepackage{bm}

\begin{document}

\title{Electronic States of Graphene Nanoribbons}

\author{L. Brey$^1$ and H. A. Fertig$^2$}

\affiliation{1. Instituto de Ciencia de Materiales de Madrid (CSIC), Cantoblanco, 28049 Madrid, Spain\\
2. Department of Physics, Indiana University, Bloomington, IN 47405}

\date{\today}

\begin{abstract}

We study the electronic states of narrow graphene ribbons (``nanoribbons'')
with zigzag and armchair edges. The finite width of these systems breaks
the spectrum into an infinite set of bands, which we demonstrate can be
quantitatively understood using the Dirac equation with appropriate
boundary conditions.  For the zigzag nanoribbon we demonstrate that the
boundary condition allows a particle- and a hole-like band with
evanescent wavefunctions confined to the surfaces, which continuously
turn into the well-known zero energy surface states as
the width gets large.
For armchair edges, we show that the boundary condition
leads to admixing of valley states, and the band structure is metallic
when the width of the sample in lattice constant units is divisible
by 3, and insulating otherwise.  A comparison of the wavefunctions
and energies from tight-binding calculations and solutions of the
Dirac equations yields quantitative agreement for all but the narrowest
ribbons.

\end{abstract}
  \maketitle

\section{Introduction}

Improvements in the processing of graphite have made possible the
isolation of two dimensional carbon sheets known as
graphene \cite{Novoselov_2004}.  The experimental  observation of
the quantum Hall effect in ribbons of this material,  with widths in the
micron \cite{Zhang_2005} or submicron \cite{Novoselov_2005} range, indicates
unambiguously the two dimensional character of the system. The
possibility of gating and further processing such graphene sheets into
multi-terminal devices has opened a new field of carbon-based
nanoelectronics, where graphene nanoribbons could be used as
connections in nanodevices.

The electronic properties of a nanometer scale carbon system
depends strongly on its size and
geometry \cite{Saito_book,Chico_1998}. This is well known in the case
of nanotubes, which are graphene sheets rolled into cylinders \cite{falkonote}.
The geometry dependence is strongly influenced by the bipartite character of the
graphene lattice. For carbon nanotubes the wrapping
direction imposes different boundary conditions on the wavefunction
in the different sublattices, which determines
whether the system is semiconducting or metallic.

\begin{figure}
  \includegraphics[clip,width=9cm]{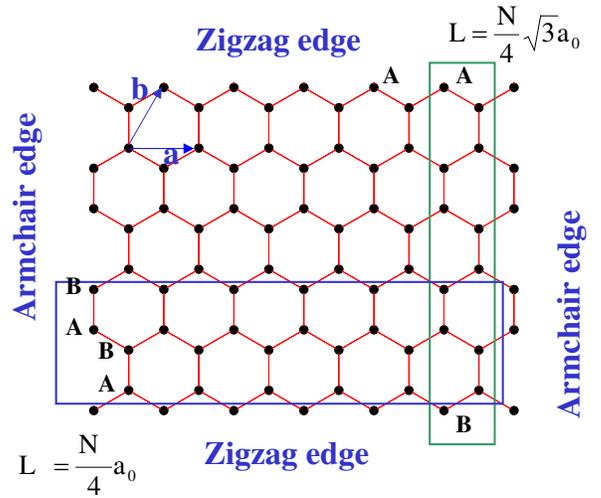}
  \caption{($Color$ $online$)The lattice structure of a graphene sheet. The primitive lattice vectors are
   denoted by ${\bf a}$ and ${\bf b}$.
Top and bottom are zigzag edges, left and right are armchair edges.
   Atoms enclosed in the vertical (horizontal) rectangle represent the unit cell used
   in the calculation of nanoribbons with zigzag (armchair) edges. The length of the nanoribbons, $L$, as function
   of the number of atoms, $N$,  in the unit cell is also indicated.}
 \label{Figure1}
\end{figure}

In this work we study the electronic states of graphene ribbons
with different atomic terminations (Fig.\ref{Figure1}). Using
tight-binding calculations, we show that the electronic properties depend
strongly on the size and geometry of the graphene nanoribbons.
We demonstrate that the electronic energies and states may be
understood in terms of eigenvalues and eigenvectors of the Dirac
Hamiltonian, which describes the physics of the electrons near
the Fermi energy of the undoped material.

We now summarize our results. We find that for nanoribbons
with zigzag edges,
the correct boundary condition is for the wavefunction
to vanish on a single sublattice at each edge. In this case the
nanoribbon has confined electronic states with wavefunctions that
involve sites on both sublattices, and is  extended across the system.
In addition, there are  surface states strongly localized near
the edges which are non-vanishing only on a single
sublattice. For armchair edges, the
appropriate boundary condition is for the wavefunction to vanish
on both sublattices at the edges. This can be achieved by admixing
states from both Dirac points. In this case we find that the
electronic structure depends critically on the nanoribbon width,
with the system being
{metallic} for nanoribbons of width  $L=3Ma_0$,
where $M$ an integer and $a_0$ the graphene lattice constant, and
insulating otherwise.

\section{Model Hamiltonian} In graphene the carbon atoms crystallize
in a honeycomb structure whose primitive lattice vectors are ${\bf a}
= a_0 (1,0) $ and ${\bf b} = a_0(1/2, \sqrt{3}/2)$. The lattice is
bipartite and there are two atoms per unit cell, denoted by A and B,
located at $(0,0)$ and at ${\bf d}= a_0(0,1/\sqrt{3})$. In the
simplest model, the carriers move in the $x$-$y$ plane by hopping between
the $p_z$ orbitals of the carbon atoms. A  tight-binding model with
only nearest neighbor hopping $t$ leads to a Hamiltonian with Dirac
points at the six corners of the Brillouin zone, only two of which
are inequivalent.  We take these to be ${\bf K}=\frac{2\pi} {a_0}
(\frac {1}{3}, \frac{1} {\sqrt{3}})$ and ${\bf
K}^{\prime}=\frac{2\pi} {a_0} (-\frac {1}{3}, \frac{1} {\sqrt{3}})$.
Wavefunctions can be expressed via the ${\bf k} \cdot {\bf P}$
approximation \cite{Ando_2005,DiVincenzo_1984} in terms of envelope
functions $[\psi_A({\bf r}),\psi_B({\bf r})]$ and
$[\psi_A^{\prime}({\bf r}),\psi_B^{\prime}({\bf r})]$ for states
near the ${\bf K}$ and ${\bf K}^{\prime}$ points, respectively,
which may be combined into a 4-vector
$\Psi=(\psi_A,\psi_B,-\psi_A^{\prime},-\psi_B^{\prime})$
\cite{Brey_Fertig_2006}. This satisfies a Dirac equation $H\Psi =
\varepsilon \Psi$, with
\begin{equation}
H=\gamma a_0\,
\left( \begin{array} {cccc} 0 & -k_x+ik_y & 0 & 0 \\
-k_x-ik_y & 0 & 0 & 0 \\
0 & 0& 0 & k_x+ik_y \\
0 & 0 & k_x-ik_y & 0 \
\end{array} \right) \, \, \, , \label{hamilt_kp}
\end{equation}
where $\gamma=\sqrt{3}t/2$.  Note that ${\bf k}$ denotes the
separation in reciprocal space of the wavefunction from the ${\bf
K}$ (${\bf K}^{\prime}$) point in the upper left (lower right) block
of the Hamiltonian.

The bulk solutions of Hamiltonian (\ref{hamilt_kp}) are well-known
\cite{Ando_2005}.
The eigenstates retain their valley index as a good quantum number
and  the  wavefunctions, with energies $\varepsilon = \pm \gamma a_0
|{\bf k}|$, may be written as $[e ^{i {\bf k} {\bf r}} e ^{- i
\theta _{{\bf k}}/2} ,\mp e ^{i {\bf k} {\bf r}} e ^{ i \theta
_{{\bf k}}/2}, 0,0 ]$ for the $\bf K$ valley , and $[0,0,e ^{i {\bf
k} {\bf r}} e ^{ i \theta _{{\bf k}}/2} ,\pm e ^{i {\bf k} {\bf r}}
e ^{- i \theta _{{\bf k}}/2} ]$ for the ${\bf K}'$ valley. Here
$\theta _{\bf k}= \arctan {k_x/k_y}$. Note that a solution to the
Dirac equation $(\psi_A,\psi_B,-\psi_A^{\prime},-\psi_B^{\prime})$
with energy $\varepsilon$ has a particle-hole conjugate partner
\cite{Ryu_2002} $(\psi_A,-\psi_B,-\psi_A^{\prime},\psi_B^{\prime})$
with energy $-\varepsilon$. Because of this, the eigenstates of
Eq.\ref{hamilt_kp} must be normalized on each sublattice separately
\cite{Brey_Fertig_2006}: $\int d{\bf r} [|\psi_{\mu}({\bf
r})|^2+|\psi_{\mu}^{\prime}({\bf r})|^2 ]= 1/2$, for $\mu=A,B$.

\begin{figure}
  \includegraphics[clip,width=9cm]{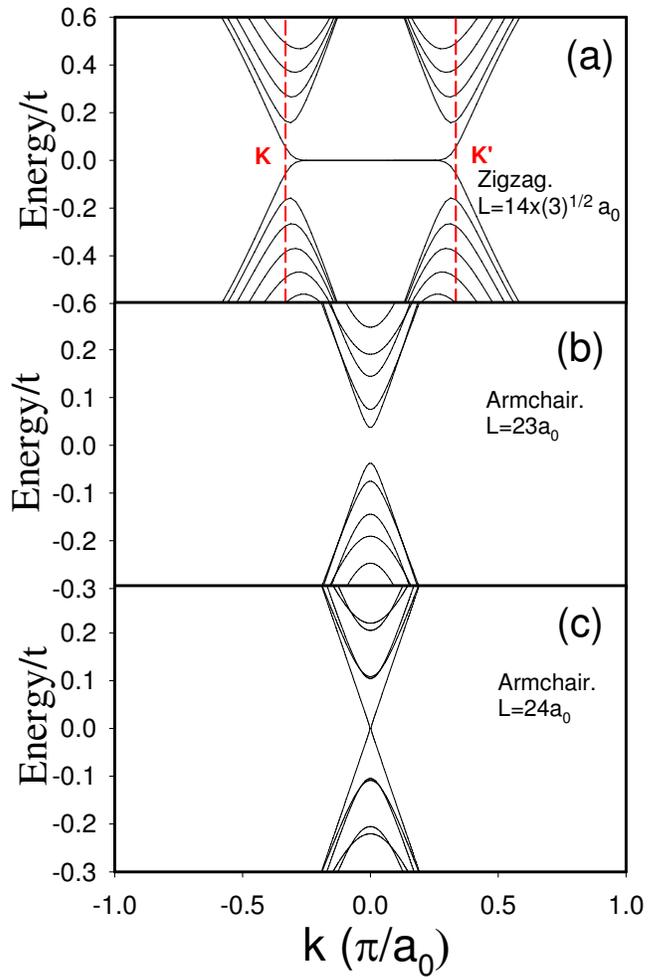}
  \caption{($Color$ $online$) Examples of energy bands for a graphene nanoribbon
  with periodic boundary conditions in one direction.
  $k$ is the wavevector parallel to the nanoribbon edge, measured with respect
  the center of the Brillouin center. (a) Ribbon terminated in zigzag edges with 56 atoms in the unit
  cell. The dispersionless states correspond to confined surface
  states. The band structures of  insulating and metallic armchair
  nanoribbons are plotted in (b) and (c) respectively.}
 \label{Figure2}
\end{figure}

\section{Zigzag Nanoribbons}
The geometry of a nanoribbon with zigzag edges is illustrated on the
top and bottom edges of Fig.~{\ref{Figure1}. It is interesting to
note that the atoms at each edge are of the same sublattice  (A on
the top edge of Fig.~{\ref{Figure1} and B on the bottom edge).
In Fig.~{\ref{Figure1} we also show the unit cell used in the
tight-binding  calculations of the zigzag ribbons, containing  $N/2$
A-type atoms that alternate  along the unit cell with $N/2$ B-type
atoms.  The total width of the nanoribbon is $L =\frac{N}{4} \,
\sqrt{3} \, a _0$. We impose  periodic boundary conditions along the
direction parallel to the edge. In our discussions we will assume
that the edges lie along the $\hat y$ direction, so in in the
discussion of the zigzag nanoribbons, the coordinate axes in
Fig.~{\ref{Figure1} will be rotated by 90$^0$, and the eigenstates
are proportional to $e^{i k y}$. In Fig. \ref{Figure2} we plot an
example of the band structure of a nanoribbon with zigzag edges. The
finite width of the ribbon produces confinement of the electronic
states near the Dirac points. In Fig.~\ref{Figure3} we plot the
energy of the first three confined states, at $k=K_y$, as a function
of the nanoribbon width. The two bands of dispersionless
localized surface states
\cite{Fujita_1996,Wakabayashi_2006,Ezawa_2006,Peres_2005} that
occur between $K_y$ and $K_y'$ in Fig. \ref{Figure2}(a) are also
affected by the finite width: they admix, and the two bands are
slightly offset from zero. The dependence of the electronic states
on the width of the nanoribbon may be understood in terms of
eigenstates of the Dirac Hamiltonian with appropriate boundary
conditions: setting the wavefunction to zero on the $A$ sublattice
on one edge, and on the $B$ sublattice for the other. We can
understand the lines of vanishing wavefunction to be lattice sites
that would lie just beyond the edges if bonds had not been cut to
form them.

\begin{figure}
  \includegraphics[clip,width=9cm]{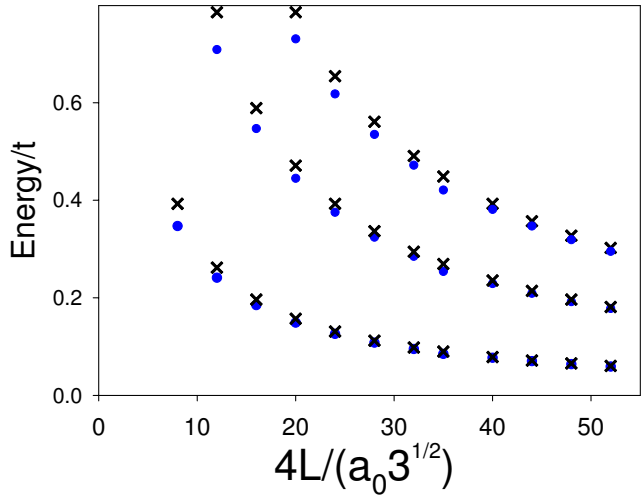}
  \caption{($Color$ $online$) Calculated confined state energies
at a Dirac point  versus the nanoribbon  width,  in
  a zigzag nanoribbon. The dots are
  tight binding results, and the crosses are
the results of the ${\bf k}\cdot{\bf P}$ approximation.}
 \label{Figure3}
\end{figure}

For the continuum description, we begin by rotating the wavevectors
in Eq. \ref{hamilt_kp}, $k_x \rightarrow k_y,~k_y \rightarrow -k_x$
so that the zigzag edge lies along the $\hat{y}$, and our
wavefunctions exist in the space $0<x<L$. Translational invariance
in the $\hat y$ direction guarantees the wavefunctions can be
written in the form $\psi_{\mu}(^{\prime})({\bf r}) = e ^{i k_y y}
\phi _{\mu}(')(x) $. For the ${\bf K}~({\bf K}')$ valley the
wavefunctions obey
\begin{eqnarray}
\left (- \partial _x ^2 + k _y ^2 \right ) \phi _B (') & =
& \tilde{\varepsilon} ^2 \phi_A(') \nonumber \\
 \left (- \partial _x ^2 + k _y ^2 \right ) \phi _A (') &
= & \tilde{\varepsilon }^2 \phi_B(') \label{Hsquare}\, \,
\end{eqnarray}
with $\tilde{\varepsilon} = {\varepsilon}  /(\gamma a_0)$. It is easy to
see if one solves the equations for $\phi _B$ and $\phi_A '$, the
remaining wavefunctions are determined by
\begin{eqnarray}
{\tilde{\varepsilon}} \, \phi_B & = &   (i \partial _x - i k _y)
\phi_A
\nonumber \\
{\tilde{\varepsilon}} \, \phi_A ' & = & (-i \partial _x + i k _y)
\phi _B ' \, \, . \label{relation}
\end{eqnarray}
The general solutions of Eq.(\ref{Hsquare}) have the form
\begin{equation}
\phi_{\mu} (x) = A e ^{zx}+Be^{-zx} \, \, ,
\end{equation}
with $z = \sqrt { k _y ^2 - \tilde{\varepsilon} ^2}$, which can be
real  or imaginary.

For the zigzag nanoribbon, we meet the boundary condition for
each type of wavefunction separately:
\begin{equation}
\phi_A
(x\!=\!0)\!=\!\phi_A'(x\!=\!0)\!=\!\phi_B(x\!=\!L)\!=\!\phi_B'(x\!=\!L)\!=\!0
\, \, .
\end{equation}
These conditions leads to a transcendental equation for
the allowed values of $z$,
\begin{equation} \frac { k _y -z } { k _y
+z} = e ^{ -2 L z} \, \, \, . \label{trans}
\end{equation}
Eq.(\ref{trans}) supports solutions with real values of $z \equiv k$
for $k_y>k _y ^c = 1/L$, which correspond to the surface states.
These have energies $\pm \sqrt{ k_y ^2 - k ^2}$, and are linear
combinations of states localized on the left and right edges of the
ribbon. For large values of $k_y$, $k \rightarrow k_y$ and the
surface states become decoupled.  For $k_y < 0$ there are no states
with real $z$ that can meet the boundary conditions, so surface
states are absent. For values of $k_y$ in the range $0<k_y<k_y^c$,
the surface states are so strongly admixed that, as we show below,
they are indistinguishable from confined states.

For pure imaginary $z=i k _n$, the transcendental equation becomes
\begin{equation}
k_y = \frac {k _n} {\tan{(k_n L)}} \, \, \, , \label{trans1}
\end{equation}
and for each solution $k_n$ there are two confined states with
energies $\tilde{\varepsilon} = \pm \sqrt{ k _n ^2 + k _y ^2}$ and
wavefunctions
\begin{equation}
\left (
\begin{array} {c}
\phi _A \\ \phi_B \end{array} \right ) = \left (
\begin{array} {c} \sin ( k _n x)
\\ \pm \frac{i}{\tilde{\varepsilon}} ( - k_n \cos (k_n x)+ k_y \sin ( k_n x) )
\end{array} \right ) \label{wf_conf} \, \, \, \, .
\end{equation}
Here the index $n$ indicates the number of nodes of the confined
wavefunction. Interestingly, for values of $k_y$ larger than $k_y
^c$, Eq.(\ref{trans1}) does not support
nodeless solutions, indicating the existence of surface states in
this region of reciprocal space. The critical value $k_y^c$ is
the momentum where the lowest energy solution of the
transcendental Eq.~\ref{trans} changes from pure real to pure
imaginary, and the energy  is equal to $\pm |k _y ^c|$.

\begin{figure}
  \includegraphics[clip,width=9cm]{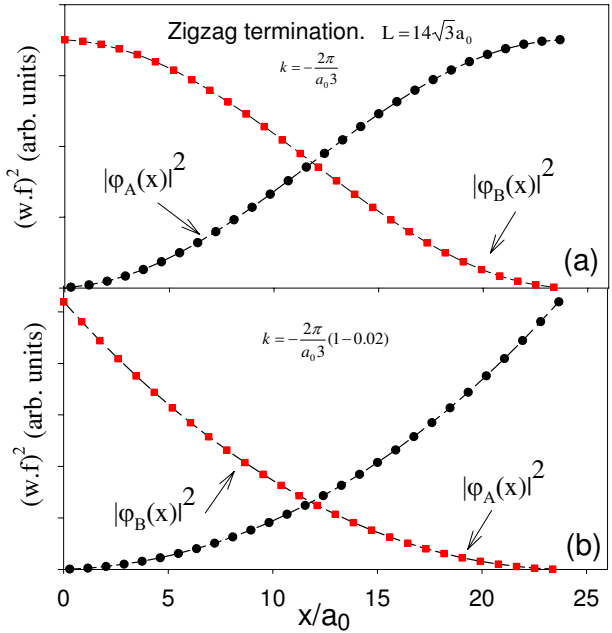}
  \caption{($Color$ $online$) Squared wavefunction for the
state closest to zero energy
  for a zigzag nanoribbon, as obtained from
  tight binding calculations.  The width of the ribbon is $L=14\sqrt{3} a_0$.
  (a)   $k= -2 \pi /3 a_0 $, and (b) $k= (-2 \pi /3 a_0) (1-0.02)
  $. Both are measured from the center of the Brillouin zone.}
 \label{Figure4}
\end{figure}

In order to analyze the accuracy of the ${\bf k} \cdot {\bf P}$
approximation for describing the electronic properties of carbon
nanoribbons,  in Fig.~\ref{Figure3} we plot the energies of the three lowest
confined states of a zigzag  nanoribbon as a function of its
width, both from the tight-binding approach and from our solutions
to the Dirac equation. It is apparent that the two approaches
match quite well, even for rather small widths ($\sim 35
\AA$).

In Fig.~\ref{Figure4} we plot the squared wavefunction for the
lowest energy state of a zigzag nanoribbon as obtained in the tight
binding approach.  Fig.~\ref{Figure4}(a) corresponds to $k_y =0$
($k= -2 \pi / 3 a_0$ with respect the center of the Brillouin zone),
and Fig.~\ref{Figure4}(b) to $k_y=0.02 \times 2 \pi / 3 a_0$. The
first case corresponds to a nodeless confined state, and we find the
wavefunction is described nearly perfectly by Eq.(\ref{wf_conf}),
whereas the second case is the expected linear combination of
surface state wavefunctions that decay exponentially from the edges
as $\exp {(-kx)}$.

\section{Armchair nanoribbons}
The geometry of a nanoribbon with armchair edges is illustrated on
the left and right edges of Fig.~{\ref{Figure1}, along with
the unit cell used in the corresponding tight binding calculations.
In this orientation the
width  of the nanoribbon is related to the number
of atoms in the unit cell through the expression $L =\frac{N}{4} \,
a _0$. Here the edge runs along the $\hat{y}$ direction, and no
rotation of the figure is needed to represent our calculations.

The electronic properties of armchair nanoribbons depend strongly
on their width. In Fig.~\ref{Figure2}(b) and (c) we plot two
examples of band structures of armchair nanoribbons.
One sees that in the latter figure there is a Dirac
point, leading to metallic behavior for a non-interacting
model, whereas the former is a band insulator.  In
general we find that armchair
nanoribbons of width $L=3Ma_0$, with $M$ integral, are metallic, whereas all the other
cases are insulators. The energy of the confined states also behave
in a discontinuous way with respect to the width of the ribbon. In
Fig.~\ref{Figure5} we plot the energy of the lowest (squared) energy confined
states at the center of the Brillouin zone as a function of the
nanoribbon width.  In the inset of this figure we see that the
separation in energy between confined states is also
strongly dependent on the number of atoms in the unit cell.

\begin{figure}
  \includegraphics[clip,width=9cm]{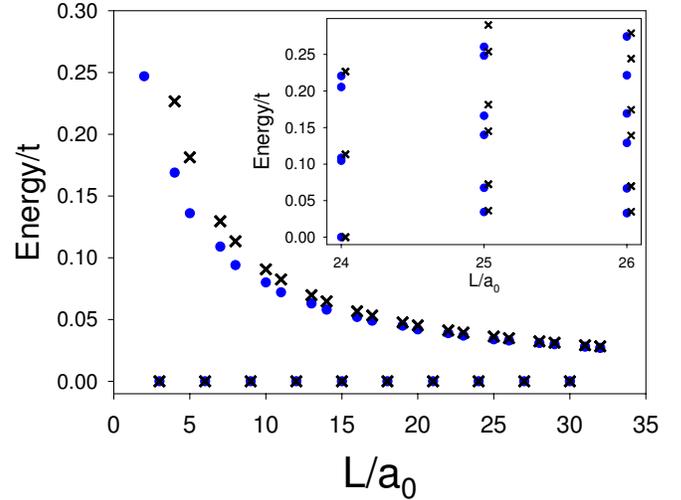}
  \caption{($Color$ $online$) Calculated lowest energy confined states at the center of
  the Brillouin zone versus the nanoribbon width,  for
  an armchair nanoribbon. The dots correspond to the
  tight binding results and the crosses are the results of the ${\bf k}\cdot{\bf P}$ approximation.
  In the inset we plot the six lowest energy confined states for three different widths.
   The ${\bf k}\cdot{\bf P}$ results are slightly shifted
  to the right for clarity. Note that the for $L=24a_0$ the ${\bf k}\cdot{\bf P}$ results are doubly degenerate. }
 \label{Figure5}
\end{figure}

As in the case of the zigzag nanoribbons this behavior may be
understood in terms of eigenstates of the Dirac Hamiltonian with
the correct boundary conditions.
In Fig.~\ref{Figure1}  one may see that
the termination consists of a line of A-B dimers, so it is natural to
have the wavefunction amplitude vanish on both sublattices at $x=0$
and $x=L$. To do this we must admix valleys, and require
\begin{eqnarray}
\phi _{\mu} (x=0) & = & \phi ' _{\mu}(x=0)  \nonumber \\
\phi _{\mu} (x=L) & = & \phi ' _{\mu} (x=L) \, e ^{i \Delta K \,
L}  \,\, , \label{bc_armchair} \nonumber
\end{eqnarray} with $\Delta K = \frac {2 \pi}{3 a_0}$.
With these
boundary conditions the general solutions of the Dirac equation
are planes waves,
\begin{equation}
\phi _B ( x )  =  e ^{i k _n x} \, \, \, \,  \textrm{and} \, \, \,
\, \phi _B '( x )  =  e ^{-i k _n x}.
\end{equation}
The wavefunctions on the A sublattice may be obtained
via
Eq.~\ref{relation}.
The wavevector $k_n$ satisfies the condition
\begin{equation}
e ^{ 2 i k _ n L} = e ^{ i \Delta K \, L} \, \, \, ,
\end{equation}
so that
\begin{equation}
2 k _n L = \frac {2  \pi }{3} \, j+ 2 \pi \, n  \, \, \, ,
\end{equation} with $n$ an integer and $j = 0,\pm 1$,
determined by
\begin{equation}
\frac {N}{4} = 3 M +j \, \, \, \,
\end{equation}
for an integer $M$. Thus for armchair nanoribbons the allowed
values of $k_n$ are
\begin{equation}
k_n = \frac {4  }{N} \pi  \left ( n \pm \frac {j}{3} \right ) \,
\, \, \, \label{condition1}
\end{equation}
with  energies $\pm \sqrt {  k _n ^2 + k _y ^2 }$. Note that this is
in contrast to the zigzag nanoribbon for which the allowed values of
$k_n$ depend on $k_y$. For a width that is a multiple of $3a_0$, the
allowed values of $k_n$,  $k_n = n 4 \pi /N$,  create doubly
degenerate states for $|n| \geq 0$, and allow a zero energy state
when $k_y \rightarrow 0$. Nanoribbons of widths that are not
multiples of three have nondegenerate states and do not include a
zero energy mode. Thus these nanoribbons are band insulators. The
quality of the ${\bf k} \cdot {\bf P}$ approximation for describing
the electronic states of armchair nanoribbons is reflected in
Fig.~\ref{Figure5} where the energies of the confined states
obtained by diagonalizing the tight binding Hamiltonian and by
solving Eq.(\ref{condition1}) are compared. The quantitative
agreement is apparent for all but the narrowest ribbons, where one
does not expect the ${\bf k} \cdot {\bf P}$ to work well.

\begin{figure}
  \includegraphics[clip,width=9cm]{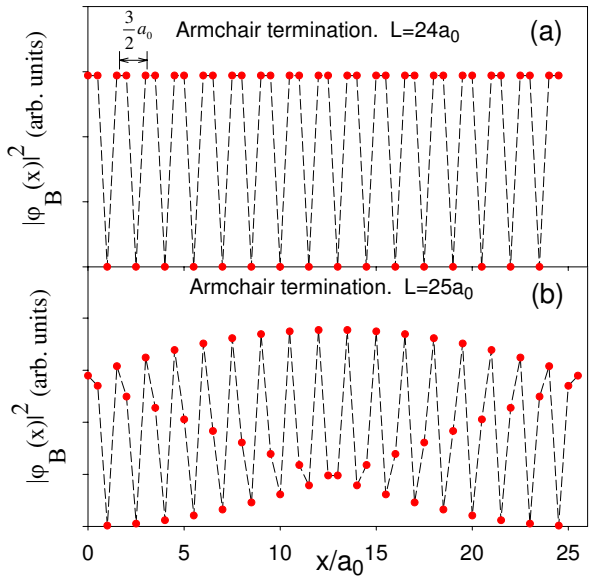}
  \caption{($Color$ $online$)
Squared wavefunction of the state  with energy closest to
zero for an armchair ribbon of width (a) $L=24 a_0$ and (b)
$L=25 a_0$, as obtained from tight binding calculations. }
 \label{Figure6}
\end{figure}

The admixing of different valley states to meet the boundary
condition means that the wavefunction will oscillate with period
$2\pi/\Delta K$ \cite{Brey_Fertig_2006}. This behavior can explicitly be seen in
Fig.~\ref{Figure6}, which illustrates the squared wavefunction
from the tight binding calculation. The short oscillation in the
wavefunctions has exactly the period expected for the valley mixing
we introduced to meet the boundary conditions. In the
case of Fig.~\ref{Figure6}(a), the ribbon is of width is $L=24 a_0$
and  $k_n =0$,  so that the energy is zero and there is no confinement
effect on the form of the wavefunction.  For a ribbon of width
$L=25 a_0$ [Fig.~\ref{Figure6}(b)], $k_n$ is non-zero and one
sees a long-wavelength oscillation whose period is related to the value of $k_n$.

\section{Conclusion}
In this paper we studied eigenstates and eigenenergies of graphene
nanoribbons using the Dirac equation with appropriate boundary
conditions, and compared the results to those of tight binding
calculations.  We found that except for the narrowest ribbons,
the agreement was quantitative.  Zigzag nanoribbons support
surface states which go to zero energy in the limit of wide
ribbons, and can only be found in a $k_y$ interval between
the Dirac points.  Armchair nanoribbons have no surface states,
but in spite of their finite size they have zero energy states
for appropriately chosen widths, so that the system oscillates
between insulating and metallic behavior as the width changes.
Our results show that the continuum description of graphene may
be used in quantitative analysis of this system for all but
the most narrow systems.

 \vspace{0.5truecm}

{\bf Acknowledgements.} The authors thank F.Guinea and C.Tejedor
for useful discussions.  This work was supported by
MAT2005-07369-C03-03 (Spain) (LB) and by the NSF through Grant No.
DMR-0454699 (HAF).


\begin{thebibliography}{14}
\expandafter\ifx\csname natexlab\endcsname\relax\def\natexlab#1{#1}\fi
\expandafter\ifx\csname bibnamefont\endcsname\relax
  \def\bibnamefont#1{#1}\fi
\expandafter\ifx\csname bibfnamefont\endcsname\relax
  \def\bibfnamefont#1{#1}\fi
\expandafter\ifx\csname citenamefont\endcsname\relax
  \def\citenamefont#1{#1}\fi
\expandafter\ifx\csname url\endcsname\relax
  \def\url#1{\texttt{#1}}\fi
\expandafter\ifx\csname urlprefix\endcsname\relax\def\urlprefix{URL }\fi
\providecommand{\bibinfo}[2]{#2}
\providecommand{\eprint}[2][]{\url{#2}}

\bibitem[{\citenamefont{K.S.Novoselov et~al.}(2004)\citenamefont{K.S.Novoselov,
  A.K.Geim, S.V.Mozorov, D.Jiang, Y.Zhang, S.V.Dubonos, I.V.Gregorieva, and
  A.A.Firsov}}]{Novoselov_2004}
\bibinfo{author}{\bibnamefont{K.S.Novoselov}},
  \bibinfo{author}{\bibnamefont{A.K.Geim}},
  \bibinfo{author}{\bibnamefont{S.V.Mozorov}},
  \bibinfo{author}{\bibnamefont{D.Jiang}},
  \bibinfo{author}{\bibnamefont{Y.Zhang}},
  \bibinfo{author}{\bibnamefont{S.V.Dubonos}},
  \bibinfo{author}{\bibnamefont{I.V.Gregorieva}}, \bibnamefont{and}
  \bibinfo{author}{\bibnamefont{A.A.Firsov}}, \bibinfo{journal}{Science}
  \textbf{\bibinfo{volume}{306}}, \bibinfo{pages}{666} (\bibinfo{year}{2004}).

\bibitem[{\citenamefont{Y.Zhang et~al.}(2005)\citenamefont{Y.Zhang, Tan,
  H.L.Stormer, and P.Kim}}]{Zhang_2005}
\bibinfo{author}{\bibnamefont{Y.Zhang}}, \bibinfo{author}{\bibfnamefont{Y.-W.}
  \bibnamefont{Tan}}, \bibinfo{author}{\bibnamefont{H.L.Stormer}},
  \bibnamefont{and} \bibinfo{author}{\bibnamefont{P.Kim}},
  \bibinfo{journal}{Nature} \textbf{\bibinfo{volume}{438}},
  \bibinfo{pages}{201} (\bibinfo{year}{2005}).

\bibitem[{\citenamefont{K.S.Novoselov et~al.}(2005)\citenamefont{K.S.Novoselov,
  A.K.Geim, S.V.Mozorov, D.Jiang, I.V.Gregorieva, S.V.Dubonos, and
  A.A.Firsov}}]{Novoselov_2005}
\bibinfo{author}{\bibnamefont{K.S.Novoselov}},
  \bibinfo{author}{\bibnamefont{A.K.Geim}},
  \bibinfo{author}{\bibnamefont{S.V.Mozorov}},
  \bibinfo{author}{\bibnamefont{D.Jiang}},
  \bibinfo{author}{\bibfnamefont{M.}~\bibnamefont{I.V.Gregorieva}},
  \bibinfo{author}{\bibnamefont{S.V.Dubonos}}, \bibnamefont{and}
  \bibinfo{author}{\bibnamefont{A.A.Firsov}}, \bibinfo{journal}{Nature}
  \textbf{\bibinfo{volume}{438}}, \bibinfo{pages}{197} (\bibinfo{year}{2005}).

\bibitem[{\citenamefont{R.Saito et~al.}(1998)\citenamefont{R.Saito,
  G.Dresselhaus, and M.S.Dresselhaus}}]{Saito_book}
\bibinfo{author}{\bibnamefont{R.Saito}},
  \bibinfo{author}{\bibnamefont{G.Dresselhaus}}, \bibnamefont{and}
  \bibinfo{author}{\bibnamefont{M.S.Dresselhaus}},
  \emph{\bibinfo{title}{Physical Properties of Carbon Nanotubes}}
  (\bibinfo{publisher}{Imperial College}, \bibinfo{address}{London},
  \bibinfo{year}{1998}).

\bibitem[{\citenamefont{L.Chico et~al.}(1998)\citenamefont{L.Chico,
  M.P.L\'opez-Sancho, and M.C.Mu$\tilde{n}$oz}}]{Chico_1998}
\bibinfo{author}{\bibnamefont{L.Chico}},
  \bibinfo{author}{\bibnamefont{M.P.L\'opez-Sancho}}, \bibnamefont{and}
  \bibinfo{author}{\bibnamefont{M.C.Mu$\tilde{n}$oz}},
  \bibinfo{journal}{Phys.Rev.Lett.} \textbf{\bibinfo{volume}{81}},
  \bibinfo{pages}{1278} (\bibinfo{year}{1998}).

\bibitem[{fal()}]{falkonote}
\bibinfo{note}{A very general discussion of boundary conditions for nanotubes
  described by the Dirac equation may be found in E. McCann and V.I. Fal'ko,
  cond-mat/0402373.}

\bibitem[{\citenamefont{T.Ando}(2005)}]{Ando_2005}
\bibinfo{author}{\bibnamefont{T.Ando}}, \bibinfo{journal}{J.Phys.Soc.Jpn.}
  \textbf{\bibinfo{volume}{74}}, \bibinfo{pages}{777} (\bibinfo{year}{2005}).

\bibitem[{\citenamefont{D.P.DiVincenzo and E.J.Mele}(1984)}]{DiVincenzo_1984}
\bibinfo{author}{\bibnamefont{D.P.DiVincenzo}} \bibnamefont{and}
  \bibinfo{author}{\bibnamefont{E.J.Mele}}, \bibinfo{journal}{Phys.\ Rev.\ B}
  \textbf{\bibinfo{volume}{29}}, \bibinfo{pages}{1685} (\bibinfo{year}{1984}).

\bibitem[{\citenamefont{L.Brey and H.A.Fertig}()}]{Brey_Fertig_2006}
\bibinfo{author}{\bibnamefont{L.Brey}} \bibnamefont{and}
  \bibinfo{author}{\bibnamefont{H.A.Fertig}}, \eprint{cond-mat/0602505}.

\bibitem[{\citenamefont{S.Ryu and Y.Hatsugai}(2002)}]{Ryu_2002}
\bibinfo{author}{\bibnamefont{S.Ryu}} \bibnamefont{and}
  \bibinfo{author}{\bibnamefont{Y.Hatsugai}}, \bibinfo{journal}{Phys.\ Rev.\
  Lett.} \textbf{\bibinfo{volume}{89}}, \bibinfo{pages}{077002}
  (\bibinfo{year}{2002}).

\bibitem[{\citenamefont{M.Fujita et~al.}(1996)\citenamefont{M.Fujita,
  Wakabayashi, K.Nakada, and K.Kusakabe}}]{Fujita_1996}
\bibinfo{author}{\bibnamefont{M.Fujita}},
  \bibinfo{author}{\bibfnamefont{K.}~\bibnamefont{Wakabayashi}},
  \bibinfo{author}{\bibnamefont{K.Nakada}}, \bibnamefont{and}
  \bibinfo{author}{\bibnamefont{K.Kusakabe}},
  \bibinfo{journal}{J.Phys.Soc.Jpn.} \textbf{\bibinfo{volume}{65}},
  \bibinfo{pages}{1920} (\bibinfo{year}{1996}).

\bibitem[{\citenamefont{K.Wakabayashi}(2006)}]{Wakabayashi_2006}
\bibinfo{author}{\bibnamefont{K.Wakabayashi}}, \emph{\bibinfo{title}{Carbon
  Bases Magnetism.}} (\bibinfo{publisher}{Elsevier}, \bibinfo{year}{2006}),
  chap. \bibinfo{chapter}{Electric and Magnetic Properties of Nanographites}.

\bibitem[{\citenamefont{M.Ezawa}(2006)}]{Ezawa_2006}
\bibinfo{author}{\bibnamefont{M.Ezawa}}, \bibinfo{journal}{Phys.Rev.B}
  \textbf{\bibinfo{volume}{73}}, \bibinfo{pages}{045432}
  (\bibinfo{year}{2006}).

\bibitem[{\citenamefont{NM.R.Peres et~al.}()\citenamefont{NM.R.Peres, F.Guinea,
  and A.H.Castro-Neto}}]{Peres_2005}
\bibinfo{author}{\bibnamefont{NM.R.Peres}},
  \bibinfo{author}{\bibnamefont{F.Guinea}}, \bibnamefont{and}
  \bibinfo{author}{\bibnamefont{A.H.Castro-Neto}}, \eprint{cond-mat/0512091}.

\end{thebibliography}

\end{document}